\documentstyle[pra,aps,preprint]{revtex} \begin{document}
  \draft
  \title{SSB of scale symmetry, fermion families \\
 and  quintessence without the long-range force problem}

\author{E. I. Guendelman \thanks{guendel@bgumail.bgu.ac.il} and
A.  B.  Kaganovich \thanks{alexk@bgumail.bgu.ac.il}} 
\address{Physics Department, Ben Gurion University of the Negev, Beer Sheva 
84105, Israel}
\maketitle
\begin{abstract} 
We study a scale invariant two measures theory where a
dilaton field $\phi$ has no explicit potentials. The scale transformations
include a translation of a dilaton $\phi\rightarrow\phi +const$. The theory
 demonstrates
a new mechanism for generation of the exponential potential: in the conformal 
Einstein frame (CEF), after SSB of scale invariance, the  theory  develops
the 
exponential potential and, in general,
 non-linear  kinetic term is generated as well.
The scale symmetry does not allow the appearance of terms breaking the
exponential shape of the potential that solves the problem of the flatness
of the scalar field potential in the context of quintessential scenarios.
As examples, two different possibilities for the choice of the
dimensionless parameters are
presented
 where the theory permits to get interesting cosmological results.
For the first choice , the theory has
standard scaling solutions for $\phi$ usually used in the context of the
quintessential scenario. For the second choice, the
theory allows three different solutions one of which is a scaling solution
with 
equation of state $p_{\phi}=w\rho_{\phi}$ where 
$w$ is predicted to be restricted by $-1<w<-0.82$. 
The regime where
the fermionic matter dominates (as compared to the dilatonic 
contribution) is analyzed. There it is found that starting from a single
fermionic field we obtain exactly three different types of spin $1/2$
particles in CEF that appears to suggest a new approach to the family
problem of particle physics. It is automatically achieved that
for two of them, fermion masses are constants, the energy-momentum tensor 
is
canonical and the "fifth force" is absent. For the third  type of
particles, a fermionic self-interaction appears as a result of SSB of
scale invariance.

   \end{abstract}          
    
    \renewcommand{\baselinestretch}{1.6} PACS number(s): 98.80.Cq, 
98.62.Gq, 11.30.Qc, 12.15.Pf

\pagebreak 

\section{Introduction} 
Recent observations imply that the Universe now is undergoing era of
acceleration\cite{P}. This is most naturally explained by the existence of
a vacuum energy which can be of the form of an explicit cosmological
constant. Alternatively, there may be a slow rolling scalar field, whose
potential (assumed to have zero asymptotic value) provides the negative
pressure required for accelerating the Universe. This is the
basic idea of the quintessence\cite{Wett1988NP668}. Some of the problems of the
quintessence scenario connected to the field theoretic grounds  of this
idea, are: 
i) what is the origin of the quintessence potential;
ii) why the asymptotic value of the potential vanishes (this is actually
the "old" cosmological constant problem\cite{CCP} );
iii) the needed flatness of the potential\cite{KLyth};
iv) without the symmetry $\phi\rightarrow\phi +const$ it is very hard to
explain the absence of the long-range force if no fine tuning is
made\cite{Carroll,Dolgov}, but such a translation-like symmetry is usually
incompatible
with a nontrivial
potential.

One of the main aims of this paper is to show how the above problems
can be solved in the
context of the two measures
theories (TMT)\cite{GK1,GK2,GK3,GK4,G,G1,G2,K}. These kind of models are based on the
observation
that in a generally covariant formulation of the action principle one has
to integrate using an invariant volume element, which is not obliged to be
dependent of the metric. In GR, the volume element $\sqrt{-g}d^{4}x$ is
indeed generally coordinate invariant, but nothing forbids us from
considering the invariant volume element $\Phi d^{4}x$ where $\Phi$ is a
scalar density that could be independent of the metric\cite{GK1}.

If the measure $\Phi$ is allowed, we have seen in a number of
models\cite{GK2,GK3,GK4,G} that, in the conformal Einstein frame (CEF),
the equations of
motion have the canonical GR structure, but the scalar field potential
produced in  the CEF is such that zero vacuum energy for the
ground
state of the theory  is obtained without fine tuning, that is the "old"
cosmological constant problem can be solved\cite{GK4}.

If  both measures ($\sqrt{-g}$ and  $\Phi$) are allowed, this opens new
possibilities concerning scale
invariance\cite{G,G1,G2,K}. In this context
we study here a theory which is invariant under
scale transformations including also a translation-like symmetry for a
dilaton field of the form $\phi\rightarrow\phi +const$ discussed by 
Carroll\cite{Carroll}. For
the case when the
original action does not contain dilaton potentials at all, it is found
that the
integration of the equation of motion corresponding to the measure $\Phi$
degrees of freedom, spontaneously breaks the scale symmetry and the
generation of a dilaton potential is a consequence of this 
spontaneous symmetry breaking (SSB). When
studying the theory in the CEF, it is demonstrated in
Sec. III that the
spontaneously induced dilaton potential has the exponential form 
and in addition, also non-linear kinetic terms appear in general. 

In Sec. IV  we discuss possible cosmological  applications of the 
theory 
when the dilaton field
is the dominant fraction of the matter: it is found that quintessential
solutions are possible.

In Sec. V we show that
in the presence of fermions, the theory displays a
successful fermionic mass generation after the spontaneous symmetries
break
(SSB), and this is actually the second main aim of this paper. In the
regime when the
fermionic density is of
the order
typical for the normal particle physics (which in the laboratory
conditions is always much higher than the dilaton density ),
 there are constant fermion masses, gravitational equations
are
canonical and the "fifth force" is absent, - all this without any
additional restrictions on the parameters of the theory.
A possible  way for explanation  to the "family puzzle" of particle
physics
also appears naturally in the context of this model. For one of the 
families, a  fermion self-interaction appears as a result of the SSB of
scale symmetry.  

\section{Two Measures Theory (TMT)}

The main idea of these kind of theories\cite{GK1,GK2,GK3,GK4} is to reconsider the
basic structure of generally relativistic actions, which are usually taken
to be of the form
\begin{equation}
    S = \int d^{4}x\sqrt{-g}L
\label{SE}
\end{equation}
where $L$ is a scalar and $g=\det(g_{\mu\nu})$. The volume element
$d^{4}x\sqrt{-g}$ is an invariant entity. It is however possible to build
a different invariant volume element if another density, that is an object
having the same transformation properties  as $\sqrt{-g}$, is introduced.
For example, given four scalar fields  
$\varphi_{a}$, $a=1,2,3,4$ we can build the density
\begin{equation}
\Phi
=\varepsilon^{\mu\nu\alpha\beta}\varepsilon_{abcd}\partial_{\mu}\varphi_{a}
\partial_{\nu}\varphi_{b}\partial_{\alpha}\varphi_{c}\partial_{\beta}\varphi_{d}
\label{Phi}
\end{equation}
and then $\Phi d^{4}x$ is also an invariant object. Notice also that
$\Phi$ is a total derivative since
\begin{equation}
\Phi
=\partial_{\mu}(\varepsilon^{\mu\nu\alpha\beta}\varepsilon_{abcd}\varphi_{a}
\partial_{\nu}\varphi_{b}\partial_{\alpha}\varphi_{c}\partial_{\beta}
\varphi_{d})
\label{Phideriv}
\end{equation}
Therefore if we consider possible actions which use both $\Phi$ and
$\sqrt{-g}$ we are lead to TMT
\begin{equation}
    S = \int L_{1}\Phi d^{4}x +\int L_{2}\sqrt{-g}d^{4}x
\label{S}
\end{equation}

Since $\Phi$ is a total derivative, a shift of $L_{1}$ by a
constant, $L_{1}\rightarrow L_{1}+const$, has the effect of adding to S
the integral of a total derivative , which does not change equations of
motion. Such a feature is not present in  the second piece of Eq.
(\ref{S})
since $\sqrt{-g}$ is not a total derivative. It is clear then that the
introduction of a new volume element has consequences on the way we think
about the cosmological constant problem, since the vacuum energy is related
 to the coupling
of the volume element with the Lagrangian. How this relation is modified
when a new volume element is introduced, was discussed in
\cite{GK2,GK3,GK4}.

It has been shown that a wide class of TMT models\cite{GK4}, containing
among others a scalar field, can be formulated which are free of the "old"
cosmological constant problem. An  important  feature of those models
consists in the use of the
"first order formalism" where the
connection coefficients $\Gamma^{\lambda}_{\mu\nu}$, metric $g_{\mu\nu}$ and 
in our case also $\varphi_{a}$  and any
matter fields that may exist are treated as independent dynamical
variables. Any relations that they satisfy are a result of the
equations of motion.
The models allow the use of the so called conformal  Einstein frame (CEF)
where the equations of motion
have canonical GR form and the effective potential
has  an absolute minimum at zero value of the effective energy density 
without fine tuning. This was verified to be the case in all examples 
studied in Ref. \cite{GK4}, provided the action form (\ref{S}) is
preserved, 
where $L_{1}$ and $L_{2}$ are $\varphi_{a}$-independent. If this is so, 
an infinite symmetry appears\cite{GK4}: 
$\varphi_{a}\rightarrow\varphi_{a}+f_{a}(L_{1})$, where $f_{a}(L_{1})$
is an arbitrary function of  $L_{1}$.

\section{Scale invariant model with spontaneous symmetry\\ breaking
 giving rise to a potential}

If we believe that there are no fundamental scales in physics, we are lead 
to the notion of scale invariance. In the context of TMT, to
implement
global scale invariance one has to introduce a "dilaton" field\cite{G,G1}.
In this case the measure $\Phi$ degrees of freedom also can participate in 
the scale transformation\cite{G,G1}. In \cite{G,G1}, explicit potentials 
(of exponential form) which respect the symmetry were introduced.
Fundamental theories however, like string theories, etc. give most naturally
only massless particles, which means that only kinetic terms and no explicit 
potentials appear from the beginning naturally. Let us  therefore explore
a similar situation in the context of a scale invariant TMT model. We postulate
then the form of the action
\begin{equation}
S=\int d^{4}x\Phi e^{\alpha\phi/M_{p}}\left[
-\frac{1}{\kappa}R(\Gamma ,g)+\frac{1}{2}g^{\mu\nu}\phi_{,\mu} 
\phi_{,\nu}\right]
+\int d^{4}x\sqrt{-g}e^{\alpha\phi/M_{p}}\left[
-\frac{b_{g}}{\kappa}R(\Gamma ,g)+\frac{b_{k}}{2}g^{\mu\nu}\phi_{,\mu} 
\phi_{,\nu}\right]
\label{totac} 
\end{equation}
where we proceed in the first order formalism and 
$R(\Gamma,g)=g^{\mu\nu}R_{\mu\nu}(\Gamma)$,\hspace{0.25cm} 
$R_{\mu\nu}(\Gamma)=R^{\alpha}_{\mu\nu\alpha}(\Gamma)$ and 
$R^{\lambda}_{\mu\nu\sigma}(\Gamma)\equiv
\Gamma^{\lambda}_{\mu\nu,\sigma}+ 
\Gamma^{\lambda}_{\alpha\sigma}\Gamma^{\alpha}_{\mu\nu}- 
(\nu\leftrightarrow\sigma)$. 
By means of a redefinition of factors of $\phi$ and of $\Phi$
one can always normalize the kinetic term of $\phi$ and the $R$-term that 
go together with $\Phi$ as done in  (\ref{totac}). Once this is done, 
this freedom however is not present any more concerning the second part of 
the action going together with $\sqrt{-g}$. The appearance of the constants
$b_{g}$ and $b_{k}$ is a result of this.

The action (\ref{totac}) is invariant under the scale transformations: 
\begin{eqnarray}
   && g_{\mu\nu}\rightarrow 
e^{\theta}g_{\mu\nu}, \quad
    \phi\rightarrow \phi-\frac{M_{p}}{\alpha}\theta ,\quad
\Gamma^{\sigma}_{\mu\nu}\rightarrow \Gamma^{\sigma}_{\mu\nu}, 
\nonumber\\ 
&&\varphi_{a}\rightarrow \lambda_{a}\varphi_{a},\quad a=1,2,3,4
\quad
where \quad \Pi\lambda_{a}=e^{2\theta}. 
\label{st} 
\end{eqnarray}

Notice that (\ref{totac}) is the most general action of TMT 
invariant under the scale transformations (\ref{st}) where the Lagrangian
densities $L_{1}$ and $L_{2}$ are linear in the scalar curvature and 
quadratic in the space-time derivatives of the dilaton but {\it without
explicit potentials}. In Refs. \cite{G,G1}, actions of such type were 
discussed, but with explicit potentials and without kinetic term going
with $\sqrt{-g}$. A different definition
of the metric have been used also in \cite{G,G1} ($g^{\mu\nu}$ in
\cite{G,G1}
instead of
the combination
$e^{\alpha\phi/M_{p}}g^{\mu\nu}$ here) so that no factor 
$e^{\alpha\phi/M_{p}}$ appeared multiplying $\Phi$ in Ref.\cite{G,G1}.
Also it is possible to formulate a consistent scale invariant model
keeping only the simplest structure (namely, only the measure $\Phi$ is
used), provided $L_{1}$ contains 4-index field strengths and an
exponential potential for the dilaton\cite{G2}. Then SSB of the scale
invariance can lead to a quintessential potential\cite{G2}. Another type
of the field theory models with explicitly broken scale symmetry have been
studied in  Ref.\cite{K} where it is shown that the quintessential
inflation\cite{PV} type models can be obtained without fine tuning.

We examine now the equations of motion that arise from (\ref{totac}).
Varying the measure fields $\varphi_{a}$, we get
\begin{equation} 
A^{\mu}_{a}\partial_{\mu}\left[ 
e^{\alpha\phi/M_{p}}\left(-\frac{1}{\kappa}R(\Gamma
,g)+
\frac{1}{2}g^{\alpha\beta}\phi_{,\alpha}\phi_{,\beta}\right)\right]=0
\label{7}
\end{equation}
\begin{equation} 
A^{\mu}_{a}=\varepsilon^{\mu\nu\alpha\beta}\varepsilon_{abcd}
\partial_{\nu}\varphi_{b}\partial_{\alpha}\varphi_{c}
\partial_{\beta}\varphi_{d}.
\label{8}
\end{equation}

Since
$Det (A^{\mu}_{a})
= \frac{4^{-4}}{4!}\Phi^{3}$ it follows that if $\Phi\neq 0$, 
\begin{equation}
e^{\alpha\phi/M_{p}}\left[ -\frac{1}{\kappa}R(\Gamma,g)+
\frac{1}{2}g^{\mu\nu}\phi_{,\mu}\phi_{,\nu}\right]=sM^{4}=const,
\label{varphi}
\end{equation}
where $s=\pm 1$ and $M$ is a constant with the dimension of mass.
It can be noticed that the appearance of a nonzero integration 
constant $sM^4$ spontaneously breaks the scale invariance (\ref{st}).

The variation of $S$ with respect to $g^{\mu\nu}$ yields
\begin{equation}
-\frac{1}{\kappa}R_{\mu\nu}(\Gamma)(\Phi +b_{g}\sqrt{-g})+
\frac{1}{2}\phi_{,\mu}\phi_{,\nu}
(\Phi +b_{k}\sqrt{-g})-
\frac{1}{2}\sqrt{-g}g_{\mu\nu}
\left[-\frac{b_{g}}{\kappa}R(\Gamma ,g)+
\frac{b_{k}}{2}g^{\alpha\beta}\phi_{,\alpha}\phi_{,\beta}\right]=0
\label{varg}
\end{equation}

Contracting Eq. (\ref{varg}) with $g^{\mu\nu}$, solving for $R(\Gamma ,g)$
and inserting into Eq. (\ref{varphi}) we obtain the constraint
\begin{equation}
M^{4}(\zeta -b_{g})e^{-\alpha\phi/M_{p}}+
\frac{\Delta}{2}g^{\alpha\beta}\phi_{,\alpha}\phi_{,\beta}
=0,
\label{con1}
\end{equation}
where the scalar $\zeta$ is the ratio of two measures
\begin{equation}
\zeta \equiv\frac{\Phi}{\sqrt{-g}}
\label{zeta}
\end{equation}
and $\Delta =b_{g}-b_{k}$. It is very interesting that {\it the geometrical
quantity $\zeta$ is defined by a constraint where neither Newton constant
nor curvature enter}.

Varying the action with respect to $\phi$ and using Eq. (\ref{varphi})
we get
\begin{equation} 
(-g)^{-1/2}\partial_{\mu}\left[(\zeta +b_{k})e^{\alpha\phi/M_{p}}
\sqrt{-g}g^{\mu\nu}\partial_{\nu}\phi)\right]-
\frac{\alpha}{M_{p}}\left[M^{4}(\zeta +b_{g})
-\frac{\Delta}{2}g^{\alpha\beta}\phi_{,\alpha}\phi_{,\beta}
e^{\alpha\phi/M_{p}}\right]=0
\label{se}
\end{equation}

Considering the term containing connection $\Gamma^{\lambda}_{\mu\nu}$, that is 
$R(\Gamma ,g)$, we see that it can be written as
\begin{equation}
S_{\Gamma}=-\frac{1}{\kappa}\int \sqrt{-g}e^{\alpha\phi/M_{p}}(\zeta +b_{g})
g^{\mu\nu}R_{\mu\nu}(\Gamma) =
-\frac{1}{\kappa}\int \sqrt{-\tilde{g}}\tilde{g}^{\mu\nu}R_{\mu\nu}(\Gamma),
\label{Gamac}
\end{equation}
where $\tilde{g}_{\mu\nu}$ is determined by the conformal transformation
\begin{equation}
\tilde{g}_{\mu\nu}=e^{\alpha\phi/M_{p}}(\zeta +b_{g})g_{\mu\nu}
\label{ct}
\end{equation}

It is clear then that the variation of $S_{\Gamma}$ with respect to $\Gamma$
will give the same result expressed in terms of $\tilde{g}_{\mu\nu}$
as in the similar GR problem in Palatini formulation. Therefore, if 
$\Gamma^{\lambda}_{\mu\nu}$ is taken to be symmetric in $\mu , \nu$, then
in terms of the metric $\tilde{g}_{\mu\nu}$, the connection coefficients
$\Gamma^{\lambda}_{\mu\nu}$ are Christoffel's connection coefficients of 
the Riemannian space-time with the metric
$\tilde{g}_{\mu\nu}$:
\begin{equation}
\Gamma^{\lambda}_{\mu\nu}=
\{
^{\lambda}_{\mu\nu}\}|_{\tilde{g}_{\mu\nu}}=\frac{1}{2}\tilde{g}^{\lambda\alpha}
(\partial_{\nu}\tilde{g}_{\alpha\mu}+\partial_{\mu}\tilde{g}_{\alpha\nu}-
\partial_{\alpha}
\tilde{g}_{\mu\nu}).
\label{Gama}
\end{equation}

So, it appears that working with $\tilde{g}_{\mu\nu}$, we recover a Riemannian 
structure  for space-time. We will refer to this as the conformal Einstein
frame (CEF). Notice that $\tilde{g}_{\mu\nu}$ is invariant under the scale 
transformations (\ref{st}) and therefore the spontaneous breaking of the global
scale symmetry (see Eq. (\ref{varphi}) and discussion after it) is reduced, in CEF, 
to the spontaneous breaking of the shift symmetry $\phi\rightarrow\phi +const$
for the dilaton field.
In this context, it is interesting to notice 
that Carroll\cite{{Carroll}} pointed
to the possible role of the shift symmetry  for a scalar field in the
resolution of the long range force problem of the quintessential scenario.

Equations (\ref{varg}) and (\ref{se}) in CEF
take the following form:
\begin{equation}
G_{\mu\nu}(\tilde{g}_{\alpha\beta})=\frac{\kappa}{2}T_{\mu\nu}^{eff}
 \label{gef}
\end{equation}
\begin{equation}
T_{\mu\nu}^{eff}=\frac{1}{2}\left(1+\frac{b_{k}}{b_{g}}\right)
\left(\phi_{,\mu}\phi_{,\nu}-K\tilde{g}_{\mu\nu}\right)-
\frac{\Delta^{2}Ke^{2\alpha\phi/M_{p}}}{2b_{g}M^{4}}
\left(\phi_{,\mu}\phi_{,\nu}-\frac{1}{2}K\tilde{g}_{\mu\nu}\right)+
\tilde{g}_{\mu\nu}\frac{sM^{4}}{4b_{g}}e^{-2\alpha\phi/M_{p}}
 \label{Tmn}
\end{equation}
\begin{equation}
N\left[(-\tilde{g})^{-1/2}\partial_{\mu}
(\sqrt{-\tilde{g}}\tilde{g}^{\mu\nu}\partial_{\nu}\phi ) +
\tilde{g}^{\alpha\beta}\partial_{\alpha}\phi\partial_{\beta}
\ln N\right]
+\frac{\alpha\Delta^{2}}{M_{p}M^{4}}K^{2}e^{2\alpha\phi/M_{p}}
-\frac{\alpha M^{4}}{M_{p}}e^{-2\alpha\phi/M_{p}}=0
\label{phief}
\end{equation}

Here 
\begin{equation}
K\equiv\frac{1}{2}\tilde{g}^{\alpha\beta}\phi_{,\alpha}
\phi_{,\beta}, \quad N\equiv
b_{g}+b_{k}-\frac{\Delta^{2}}{M^{4}}Ke^{2\alpha\phi/M_{p}},
\label{not}
\end{equation}
$G_{\mu\nu}(\tilde{g}_{\alpha\beta})$
is the Einstein tensor in the Riemannian space-time with metric
$\tilde{g}_{\mu\nu}$ and  the constraint (\ref{con1}) have been used which 
in  CEF
takes the form
\begin{equation}
\zeta =b_{g}\frac{M^{4}-\Delta Ke^{2\alpha\phi/M_{p}}}
{M^{4}+\Delta Ke^{2\alpha\phi/M_{p}}}
\label{con2}
\end{equation}

Notice that in $T_{\mu\nu}^{eff}$ we can recognize an effective potential 
\begin{equation}
V_{eff}=\frac{sM^{4}}{4b_{g}}e^{-2\alpha\phi/M_{p}}
\label{Veff}
\end{equation}
 which appears in spite
of the fact that no explicit potential term was introduced in the original 
action
(\ref{totac}). As we see, the existence of $V_{eff}$ is associated with 
the constant $sM^{4}$, appearance of which spontaneously  breaks the scale 
invariance. This is actually a new mechanism for generating the
exponential potential\footnote{See for comparison Refs.
\cite{exppot1,exppot2,exppot3}
and a general discussion in Ref. \cite{FJ}}. 

Notice also that if $b_{g}\neq b_{k}$, the effective energy-momentum 
$T_{\mu\nu}^{eff}$ as well as the dilaton equation of motion contain
 the non-canonical terms nonlinear\footnote{Other possible  origin for
the non-linear kinetic terms, known in the literature\cite{WP},
 are higher order gravitational corrections in
string and supergravity theories.} in gradients of the dilaton $\phi$. 
It will be very important that the 
non-canonical in $\phi_{,\alpha}$ terms are multiplied by a very 
specific exponential of $\phi$. As we will see, these non-canonical terms 
may be responsible for the most interesting scaling solutions.
In the context of FRW cosmology, this structure
provides conditions for quintessential solutions if $s=1$.

\section{Scaling solutions}

In the context of a spatially flat FRW cosmology with a metric $ds^{2}_{eff}=
\tilde{g}_{\mu\nu}dx^{\mu}dx^{\nu} =dt^{2}-a^{2}(t)(dx^{2}+dy^{2}+dz^{2})$,
the equations (\ref{gef})-(\ref{phief}), with the choice $s=+1$, become:
\begin{equation}
H^{2}=\frac{1}{3M_{p}^{2}}\rho_{eff}(\phi)
\label{FRWgr}
\end{equation}
\begin{eqnarray}
&&\left(b_{g}+b_{k}-
\frac{\Delta^{2}}{2M^{4}}\dot{\phi}^{2}e^{2\alpha\phi/M_{p}}\right)
\left[\ddot{\phi}+3H\dot{\phi}+
\dot{\phi}\partial_{t}\ln{\Big |}
b_{g}+b_{k}-
\frac{\Delta^{2}}{2M^{4}}\dot{\phi}^{2}e^{2\alpha\phi/M_{p}}{\Big |}
\right]
\nonumber\\
&&+\frac{\alpha\Delta^{2}}{4M^{4}M_{p}}\dot{\phi}^{4}e^{2\alpha\phi/M_{p}}
-\frac{\alpha M^{4}}{M_{p}}e^{-2\alpha\phi/M_{p}}=0
\label{FRWphi}
\end{eqnarray}
where the energy density of the dilaton field is
\begin{equation}
\rho_{eff}(\phi)=\frac{1}{4}\left(1+\frac{b_{k}}{b_{g}}\right)\dot{\phi}^{2}
-\frac{3\Delta^{2}}{16b_{g}M^{4}}\dot{\phi}^{4}e^{2\alpha\phi/M_{p}}+ 
\frac{M^{4}}{4b_{g}}e^{-2\alpha\phi/M_{p}}
\label{rho}
\end{equation}
and the pressure
\begin{equation}
p_{eff}(\phi)=\frac{1}{4}\left(1+\frac{b_{k}}{b_{g}}\right)\dot{\phi}^{2}
-\frac{\Delta^{2}}{16b_{g}M^{4}}\dot{\phi}^{4}e^{2\alpha\phi/M_{p}}-
 \frac{M^{4}}{4b_{g}}e^{-2\alpha\phi/M_{p}}
\label{p}
\end{equation}

One can see that Eqs. (\ref{FRWgr})-(\ref{rho}) allow  solutions
of a familiar quintessential form\cite{Wett1988NP668,FJ}
\begin{equation}
\phi(t)=\frac{M_{p}}{2\alpha}\phi_{0}+\frac{M_{p}}{\alpha}\ln (M_{p}t)
\label{phiq}
\end{equation}
\begin{equation}
a(t)=t^{\gamma}
\label{aq}
\end{equation}
which provides scaling behaviors of the dilaton energy density
\begin{equation}
\rho_{eff}(\phi)\propto 1/a^{n}.
\label{scalrho}
\end{equation}
The important role for possibility of such solutions belongs to
the remarkable feature of the nonlinear terms in 
Eqs. (\ref{FRWgr})-(\ref{rho})  that appear only in the combination
$\dot{\phi}^{2}e^{2\alpha\phi/M_{p}}$ which remains constant for
the solutions (\ref{phiq}) and (\ref{aq}): 
\begin{equation}
\dot{\phi}^{2}e^{2\alpha\phi/M_{p}}=const
\label{constnonlinear}
\end{equation}

Eqs. (\ref{phiq})-(\ref{scalrho}) describe solutions of 
Eqs. (\ref{FRWgr})-(\ref{rho}) with $n=\frac{2}{\gamma}$ if
\begin{equation}
\gamma =\frac{b_{g}+b_{k}-y}{4b_{g}\alpha^{2}}
\label{gam}
\end{equation}
where 
\begin{equation}
y\equiv\frac{\Delta^{2}M_{p}^{4}e^{\phi_{0}}}{2M^{4}\alpha^{2}} 
\label{y}  
\end{equation}
is a solution of the cubic equation
\begin{equation}
y^{3}-2(b_{g}+b_{k}-b_{g}\alpha^{2})y^{2}+(b_{g}+b_{k})(b_{g}+b_{k}-
\frac{4}{3}b_{g}\alpha^{2})y-\frac{2}{3}b_{g}\alpha^{2}\Delta^{2}=0. 
\label{eqy}  
\end{equation}

Up to now we did not make any assumptions about parameters of the
theory. We will now suppose that $b_{g}$ and $b_{k}$ are positive and 
consider two particular cases.

{\it The case I}. If 
\begin{equation}
b_{k}=b_{g}=b
\label{keqb}
\end{equation}
then one can  immediately   see that 
Eqs. (\ref{FRWgr})-(\ref{p}) describe the FRW cosmological model
 in the context of the standard GR when the minimally coupled
scalar field $\phi$ with the
potential $\frac{M^{4}}{4b}e^{-2\alpha\phi/M_{p}}$ is the only
source of gravity. In this case the scaling solution (\ref{phiq}),
(\ref{aq}) coincides with the standard one\cite{FJ} where 
\begin{equation}
\gamma =\frac{1}{2\alpha^{2}}, \qquad n=4\alpha^{2}.
\label{qstand}
\end{equation}

{\it The case II}. Another interesting possibility consists of the
assumption that
\begin{equation}
b_{k}\ll b_{g}
\label{bkey}  
\end{equation}
Then ignoring corrections of the order of $b_{k}/b_{g}$,
the solutions  of Eq. (\ref{eqy}) are
\begin{equation}
y_{1}=b_{g}
\label{soly1}  
\end{equation}
 \begin{equation}
y_{2}=\frac{b_{g}}{2}\left[1-2\alpha^{2}+\sqrt{4\alpha^{4}-
\frac{20}{3}\alpha^{2}+1}\right]
\label{soly2}  
\end{equation}
\begin{equation}
y_{3}=\frac{b_{g}}{2}\left[1-2\alpha^{2}-\sqrt{4\alpha^{4}-
\frac{20}{3}\alpha^{2}+1}\right]
\label{soly3}  
\end{equation}

The solution $y_{1}$ corresponds to the static universe 
($\gamma =0$ and $a(t)=const$) supported by the slow rolling scalar
field $\phi$, Eq. (\ref{phiq}). However, taking into account
corrections of the order $b_{k}/b_{g}$ to $y_{1}$ we will get
$\gamma\propto {\cal O}(b_{k}/b_{g})$.

Solutions $y_{2}$ and $y_{3}$ exist and are positive (see the 
definition (\ref{y})) only if
\begin{equation}
\alpha^{2}\leq \frac{1}{6}
\label{alpha}
\end{equation}

The solution $y_2$ corresponds to the values of the parameter
$\gamma$ monotonically varying from $\gamma_{min}=2/3$
up to $\gamma =1$ as $\alpha^2$ changes from 0 up to $1/6$.

The most interesting solution is given by $y_{3}$ that provides
the values of the parameter
$\gamma$ monotonically varying from $\gamma_{min}= 1$
up to $\infty$ as $\alpha^2$ changes from $1/6$ up to zero.
In this case, Eqs. (\ref{phiq})-(\ref{aq}) describe an
accelerated  universe for all permissible values of $\alpha^2$ 
and the energy density of the dilaton field scales as in Eq. 
(\ref{scalrho}) with monotonically varying $n$, $2\geq n\geq 0$ as
 $\alpha^2$ changes from $1/6$ up to zero. For the dilatonic matter
equation-of-state $p=w\rho$ we get 
\begin{equation}
-1\leq w\leq -32/39\approx -0.82
\label{w}
\end{equation}

In the conclusion of this section let us revert to one of the problems of
the quintessence discussed in Introduction, namely to the flatness
problem\cite{KLyth}. This is a question of the field theoretic
basis
for the choice of the flat enough potential. In fact, Kolda and Lyth noted
\cite{KLyth} that an extreme fine tuning is needed in order to prevent the
contribution from another possible terms breaking the flatness of the
potential (see also for a review by Binetruy in Ref.\cite{CCP}). In the
theory we
study here,
there is a symmetry (scale
symmetry (\ref{st})) which forbids the appearance of such dangerous
contributions into $V_{eff}$, at least on the classical level.  One can
hope that the soft breaking of the scale symmetry guaranties that the
symmetry breaking quantum corrections to the classical effective 
potential (\ref{Veff}) will be small.

Here we have to make a note concerning quantization of the dilaton field. 
If $\Delta\neq 0$ then one can see from Eq. (\ref{rho}) that there is  
a possibility of negative energy contribution from the space-time 
derivatives of the dilaton. This raises of course the suspicion that the
quantum theory may contain ghosts. 
Let us see that  this problem does not appear when considering small 
perturbations around the background determined  by the  studied above
scaling solutions.
To see this, let us calculate the canonically conjugate momenta to $\phi$,
starting from the original action (\ref{totac}) and expressing it in
terms of 
the variables defined in CEF, Eq. (\ref{ct}):
\begin{equation}
\pi_{\phi}=\frac{1}{2b_{g}}\left(b_{g}+b_{k}-
\frac{\Delta^{2}}{sM^{4}}Ke^{2\alpha\phi/M_{p}}\right)\sqrt{-\tilde{g}}
\tilde{g}^{00}\dot{\phi}
\label{canconj}
\end{equation} 

As we have seen, the cosmological scaling solutions provide backgrounds 
where 
$Ke^{2\alpha\phi/M_{p}}=const$. Moreover, it is easy to see that 
for the scaling solutions
\begin{equation}
\pi_{\phi}=\frac{1}{2b_{g}}\left(b_{g}+b_{k}-y\right)a^{3}\dot{\phi}=
2\alpha^{2}\gamma a^{3}\dot{\phi},
\label{canconjscaling}
\end{equation} 
where $\gamma$ and $y$ are defined by Eqs. (\ref{gam}) and (\ref{y}).
We have seen also that for studied scaling solutions, 
 $\gamma$ gets positive values. Therefore we conclude that in such
backgrounds $\pi_{\phi}$ and $\dot{\phi}$ have the same sign, that 
guaranties a ghost-free quantization.
 The only exclusion is the particular case when $b_{k}=0$, $y=b_{g}$. As
we
have seen, such solution describes a static universe. In this case
the canonically conjugate momenta $\pi_{\phi}=0$ and therefore it appears
that in this vacuum there are no particles associated with the scalar
field $\phi$.

\section{Scale invariant fermion-dilaton coupling \\
 without the long-range
force problem}

In general scalar-tensor theories, particle masses depend on time, when
the theory is studied in the frame where Newton's constant is really a
constant. However, for all the fermionic matter observed in the universe,
the cosmological variation of particle masses (including those of electrons)
is highly constrained. We want to show now how the theory presented in 
this paper avoids this problem and also the so called fifth force problem,
in spite of the need to include exponential couplings of the dilaton field
 to fermionic matter  in order to ensure global scale invariance.

To describe fermions, normally one uses the vierbein ($e_{a}^{\mu}$) and 
spin-connection ($\omega_{\mu}^{ab}$) formalism where the metric is given by 
$g^{\mu\nu}=e^{\mu}_{a}e^{\nu}_{b}\eta^{ab}$ and the scalar curvature is
$R(\omega ,e) =e^{a\mu}e^{b\nu}R_{\mu\nu ab}(\omega)$ where
\begin{equation} 
R_{\mu\nu ab}(\omega)=\partial_{\mu}\omega_{\nu ab}
+\omega_{\mu a}^{c}\omega_{\nu cb}
-(\mu\leftrightarrow\nu).
        \label{B}
\end{equation}

Following the general idea of the model, we now treat the geometrical 
objects $e_{a}^{\mu}$, $\omega_{\mu}^{ab}$, the measure fields $\varphi_{a}$,
as well as the dilaton $\phi$ and the fermionic fields as independent variables.
In this formalism, the natural generalization of the action (\ref{totac})
keeping the general structure (\ref{S}), 
when a fermion field $\Psi$ is also present and which also respect scale 
invariance is the following:
\begin{eqnarray}
&S=& \int d^{4}x e^{\alpha\phi /M_{p}}
(\Phi +b\sqrt{-g})\left[-\frac{1}{\kappa}R(\omega ,e)
+\frac{1}{2}g^{\mu\nu}\phi_{,\mu}\phi_{,\nu}\right]
\nonumber\\
&+&\int d^{4}x e^{\alpha\phi /M_{p}}
\left[(\Phi +k\sqrt{-g})\frac{i}{2}\overline{\Psi}
\left(\gamma^{a}e_{a}^{\mu}\overrightarrow{\nabla}_{\mu}
-\overleftarrow{\nabla}_{\mu}\gamma^{a}e_{a}^{\mu}\right)\Psi 
-(\Phi +h\sqrt{-g})e^{\frac{1}{2}\alpha\phi/M_{p}}m\overline{\Psi}\Psi
\right]
\label{totaction}
\end{eqnarray}
where $\overrightarrow{\nabla}_{\mu}=\overrightarrow{\partial}_{\mu}+
\frac{1}{2}\omega_{\mu}^{cd}\sigma_{cd}$ and
$\overleftarrow{\nabla}_{\mu}=\overleftarrow{\partial}_{\mu}-
\frac{1}{2}\omega_{\mu}^{cd}\sigma_{cd}$.

The action (\ref{totaction}) is invariant
under 
the global scale transformations
\begin{eqnarray}
    e_{\mu}^{a}\rightarrow e^{\theta /2}e_{\mu}^{a}, \quad
\omega^{\mu}_{ab}\rightarrow \omega^{\mu}_{ab}, \quad
\varphi_{a}\rightarrow \lambda_{a}\varphi_{a}\quad
where \quad \Pi\lambda_{a}=e^{2\theta} 
\nonumber
\\ 
\phi\rightarrow \phi-\frac{M_{p}}{\alpha}\theta ,\quad
\Psi\rightarrow e^{-\theta /4}\Psi, \quad 
\overline{\Psi}\rightarrow  e^{-\theta /4} \overline{\Psi}. 
\label{stferm} 
\end{eqnarray}

In (\ref{totaction}) two types of fermionic "kinetic-like terms" (as well
as "mass-like terms") which 
respect scale invariance have been introduced: they are coupled to the
measure $\Phi$ and to the measure $\sqrt{-g}$ respectively.
As we have discussed in the previous section, the quantum theory may in
general contain ghosts if $b_{g}\neq b_{k}$. Taking this into account and
also for the sake of a simplification of the presentation of the results 
we have chosen $b_{g}= b_{k}= b$. Notice however that in the framework of
the classical theory, all conclusions will be made below are true also if
$b_{g}\neq b_{k}$. Except for this, Eq.(\ref{totaction}) describes the
most general action\footnote{Recall that in this paper we restrict
ourselves
to the models without explicit dilaton potentials in the original action}
 satisfying the formulated above symmetries.

We can immediately obtain the equations of motion. From these going
through
similar steps to those performed in Sec. III, a constraint follows again
which replaces (\ref{con1}) and which contains now a contribution from the
fermions. The spin-connection can be found by the variation of
$\omega^{\mu}_{ab}$. 

Similar to what we learned from the treatment of Sec.III, we can consider
the theory in the CEF which in this case involves
also a transformation of the fermionic fields:
\begin{eqnarray}
\tilde{g}_{\mu\nu}=e^{\alpha\phi/M_{p}}(\zeta +b)g_{\mu\nu}, \quad  
\tilde{e}_{a\mu}=e^{\frac{1}{2}\alpha\phi/M_{p}}(\zeta
+b)^{1/2}e_{a\mu},
\nonumber
\\
\Psi^{\prime}=e^{-\frac{1}{4}\alpha\phi/M_{p}}
\frac{(\zeta +k)^{1/2}}{(\zeta +b)^{3/4}}\Psi
\label{ctferm}
\end{eqnarray}

In  terms of these variables, the transformed spin-connections
$\tilde{\omega}_{\mu }^{cd}$ turns out to be that of the
Einstein-Cartan space-time and, besides, the new
variables $\tilde{g}_{\mu\nu}$, $\tilde{e}_{a\mu}$,
$\Psi^{\prime}$ and $\overline{\Psi}^{\prime}$ are invariant under
the scale transformations (\ref{stferm}). In the CEF 
the only field which still has a non trivial transformation property is
the dilaton $\phi$ which gets shifted (according to (\ref{stferm})).
Thus, the presence of fermions does not change a conclusion made in Sec.III
after Eq.(\ref{Gama}): the spontaneous breaking of the 
scale symmetry  is reduced, in the CEF, 
to the spontaneous breaking of the shift symmetry $\phi\rightarrow\phi +const$
for the dilaton field.

In terms of  $\tilde{e}_{a\mu}$,
$\Psi^{\prime}$, $\overline{\Psi}^{\prime}$ and $\phi$, the constraint 
(again arising as a self-consistency condition of equations of
motion) which
now replaces (\ref{con2}) and which contains now a contribution from
 the fermions is
\begin{equation}
(\zeta -b)M^{4}e^{-2\alpha\phi/M_{p}}+
F(\zeta)(\zeta +b)^{2}
m\overline{\Psi}^{\prime}\Psi^{\prime}=0.
\label{confermEin}
\end{equation}
where we have chosen $s=+1$ for definiteness and the function $F(\zeta)$ is
defined by
\begin{equation}
F(\zeta)\equiv
\frac{1}{2(\zeta +k)^{2}(\zeta +b)^{1/2}}
[\zeta^{2}+(3h-k)\zeta +2b(h-k)+kh] 
\label{F}
\end{equation}

The dilaton field equation is 
\begin{equation}
(-\tilde{g})^{-1/2}\partial_{\mu}
\left(\sqrt{-\tilde{g}}\tilde{g}^{\mu\nu}\partial_{\nu}\phi \right)
-\frac{\alpha M^{4}}{M_{p}(\zeta +b)}e^{-2\alpha\phi/M_{p}}+
\frac{\alpha m}{M_{p}}F(\zeta)\overline{\Psi}^{\prime}\Psi^{\prime}
=0.
\label{phief+ferm}   
\end{equation}

The fermionic equation of motion
in terms of the
variables (\ref{ctferm}) takes the standard structure of that 
in the Einstein-Cartan space-time\cite{Hehl}
 where a fermion
field is the only source of a non-riemannian  part of the connection. 
The only novelty of the fermionic equation consists of the
form of
the $\zeta$- depending fermion "mass" $m^{(eff)}(\zeta)$:
\begin{equation}
m^{(eff)}(\zeta)=
\frac{m(\zeta +h)}{(\zeta +k)(\zeta +b)^{1/2}}
 \label{muferm}
\end{equation}

The gravitational equations are of the standard form (\ref{gef})
with 
\begin{equation}
T_{\mu\nu}^{eff}=
\phi_{,\mu}\phi_{,\nu}-K\tilde{g}_{\mu\nu}
+\frac{b_{g}M^{4}}{(\zeta +b)^{2}}
e^{-2\alpha\phi/M_{p}}\tilde{g}_{\mu\nu}
+T_{\mu\nu}^{(f,canonical)}
-mF(\zeta)
\overline{\Psi}^{\prime}\Psi^{\prime}\tilde{g}_{\mu\nu},
 \label{Tmn+f}
\end{equation}
where 
\begin{equation}
T_{\mu\nu}^{(f,canonical)}=
\frac{i}{2}[\overline{\Psi}^{\prime}\gamma^{a}e_{a(\mu }^{\prime}
\nabla_{\nu )}\Psi^{\prime}-(\nabla_{(\mu}\overline{\Psi}^{\prime})
\gamma^{a}e_{\nu )a}^{\prime}\Psi^{\prime}]
\label{Tmnfcanon}
\end{equation}
is the canonical energy-momentum tensor for the fermionic field
in the curved space-time\cite{Birrel} and
 $\nabla_{\mu}\Psi^{\prime}
=\left(\partial_{\mu}+
\frac{1}{2}\tilde{\omega}_{\mu }^{cd}\sigma_{cd}\right)\Psi^{\prime}$
and $\nabla_{\mu}\overline{\Psi}^{\prime}=
\partial_{\mu}\overline{\Psi}^{\prime}-
\frac{1}{2}\tilde{\omega}_{\mu}^{cd}\overline{\Psi}^{\prime}\sigma_{cd}$.

The scalar field $\zeta$ is defined by the constraint (\ref{confermEin}) in
terms of the dilaton and fermion fields as a solution of the seventh
degree
algebraic equation that makes finding $\zeta$ in general a very
complicated
question. However there are two physically most interesting limiting cases
when   
solving (\ref{confermEin}) is simple enough. 

Let us first  analyze the constraint (\ref{confermEin})  
when  the fermionic  density (proportional to
$\overline{\Psi}^{\prime}\Psi^{\prime}$) is very low as compared to the 
contributions of the dilaton potential 
($\propto M^{4}e^{-2\alpha\phi/M_{p}}$).
In this limiting case,
 the constraint gives again the expression (\ref{con2})
for $\zeta$ where we have to take now $\Delta =0$,  that is constraint
yields the constant value\footnote{ Notice that if we had chosen
 $b_{g}\neq b_{k}$ and
 assumed  the quintessential cosmological solution
of Sec.IV where
$Ke^{2\alpha\phi/M_{p}}=const$, we would again get   a constant value
of $\zeta$.} $\zeta =b$.
Inserting this value of $\zeta$ into (\ref{muferm}) we see that the mass 
of a "test" fermion (that is when we ignore the effect of the fermion 
itself on the dilatonic background) is constant.

An opposite regime is realized  when the contribution of the fermionic  
density to the constraint (\ref{confermEin})  is very high
as compared to the contribution of the dilaton potential.
In the context of the present day universe,
this regime corresponds in particular to the normal laboratory conditions
in particle physics. 
 Then according to the
 constraint (\ref{confermEin}), one of the possibilities for this to be 
realized consists in the condition
\begin{equation}
F(\zeta)= 0
\label{F=0}
\end{equation}
from which we find two possible {\it constant} values for $\zeta$
\begin{equation}
\zeta_{1,2}=\frac{1}{2}\left[k-3h\pm\sqrt{(k-3h)^{2}+
8b(k-h)
-4kh}\,\right] 
\label{zeta12}
\end{equation}
These solutions, i.e. values $\zeta_{1}$ and $\zeta_{2}$,  are real
and different for very broad range of   the parameters
$b$, $k$ and $h$. These conditions have to be  considered together with
the obvious requirement $\zeta +b >0$ (see transformations
(\ref{ctferm})).
For
instance,  for $h>0$,  all these conditions are satisfied 
provided that parameters are situated in the broad region defined by the
system of inequalities $(b-h)(b-k)>0$ and
$(k-h)[k-h+8(b-h)]>0$. 

We see from (\ref{muferm}) that two different constants $\zeta$ given by
(\ref{zeta12})
define in general {\it two specific masses} for the fermion. We will
assume that these two fermionic states should be identified with the
first two fermionic generations.

The separate possibility relevant to the high fermionic  density 
(again, as compared to the contributions of the dilaton potential)
 is the case when 
\begin{equation}
\zeta +b\approx 0
\label{zeta=0}
\end{equation}
is a solution. However, the solution $\zeta +b=0$ is singular one as
 we see from equations 
of motion. This means that one can not neglect the first  term 
in the constraint (\ref{confermEin}) and 
instead of $\zeta +b=0$ we have to take the solution $\zeta_{3} \approx -b$
by solving $\zeta +b$ in terms of the dilaton field and the primordial
fermion field itself. Then it follows from (\ref{confermEin}) and
(\ref{F}) that
\begin{equation}
\frac{1}{\sqrt{\zeta_{3} +b}}\approx 
\left[\frac{m(h-b)}{4M^{4}b(k-b)}
\overline{\Psi}^{\prime}\Psi^{\prime}e^{2\alpha\phi /M_{p}}\right]^{1/3}.
\label{srtzeta}
\end{equation}
Therefore, instead of constant masses, as it was for $\zeta_{1}$ and
$\zeta_{2}$ (i.e. in the case $F(\zeta)=0$),
this leads to higher 
fermion self-interaction which can be
represented by  the following term in the effective fermion Lagrangian in
the
dilatonic background $\phi =\bar{\phi}$:
\begin{equation}
L^{ferm}_{selfint}=
3\left[\frac{1}{b}\left(\frac{m(h-b)}{4M(k-b)}
\overline{\Psi}^{\prime}\Psi^{\prime}\right)^{4}
e^{2\alpha\bar{\phi} /M_{p}}\right]^{1/3}.
\label{selfint}
\end{equation}
The coupling
constant of this self-interaction depends on the dilaton
$\phi$.
The condition (\ref{zeta=0}) is realized, for example, as the classical
cosmological background value
 $\phi =\bar{\phi}(t)\rightarrow\infty$ that corresponds to the late
universe in 
the quintessence scenario. 
A full
treatment of the case with $\zeta =\zeta_{3}$, which we assume corresponds
to the third fermion generation, 
requires the study of
quantum corrections.
We expect that after 
$\overline{\Psi}^{\prime}\Psi^{\prime}$ develops an expectation value,
the fermion condensate  will give the third family
appropriate masses
similar to what we know in NJL model
\cite{NJL} (for recent
progress in this subject see e. g. Ref. \cite{Cv}). It is interesting to
note that appearance of the higher fermion self-interaction here is
related to the
SSB of the scale invariance. In fact, the appearance of the integration
constant $M$ in Eqs. (\ref{srtzeta}) and (\ref{selfint}) tells us that
without SSB of
scale invariance such  interaction is not defined.

Concluding this analysis of equations  when  the fermionic density is of 
the order
typical for the normal particle physics (which in the laboratory 
conditions is always much higher than the dilaton density )
we see
that starting from a single primordial
fermionic field we obtain  exactly three different types
of spin $1/2$
particles in the CEF. This appears to be a new approach to the family
problem
in particle physics and it will be subject of a detail study in another
publication. 

Coming back to the first two fermion families  generated in the regime of
fermion dominance as $F(\zeta)=0$ we note that
surprisingly  the same factor $F(\zeta)$ appears in the last terms of 
Eqs. (\ref{phief+ferm}) and (\ref{Tmn+f}). Therefore, in the regime 
where {\it regular} fermionic  matter (i.e. $u$ and $d$
quarks, $e^{-}$ and $\nu_{e}$) is a dominant
fraction,
the last terms of 
Eqs. (\ref{phief+ferm}) and (\ref{Tmn+f}) {\it automatically} vanish.
In Eq. (\ref{phief+ferm}), this means that the fermion density
 $\overline{\Psi}^{\prime}\Psi^{\prime}$ is not a source for the dilaton 
and thus the long-range force disappears automatically. Notice that 
there is no need to require no interactions of the dilaton with regular
matter at all to have agreement with observations but  it is rather enough
that these interactions vanish in the regime where regular
fermionic
matter dominates over other matter fields. In Eq. (\ref{Tmn+f}), the condition
(\ref{F=0})
means that in the  region where the regular fermionic matter dominates,
 the fermion energy-momentum
tensor becomes equal to the canonical 
energy-momentum tensor of a fermion field in GR.  
\footnote{The decoupling of the dilaton in the CEF in the case of high 
fermion density was discussed also in a simpler spontaneously broken scale
invariant model
(with $b=k=0$ and explicit exponential potentials) in Ref. 
\cite{G1}. In the framework of other  TMT  model\cite{K} (with small {\it
explicit}
breaking of the scale invariance) in the
context of the quintessential scenario, a special tuning of the parameters
is needed to achieve the dilaton-fermion decoupling.}

\section{Discussion and conclusions}

In this paper the possibility of a spontaneously generating exponential
potential for the dilaton field in the context of TMT with spontaneously
broken global scale symmetry was studied. The symmetry transformations
formulated
in terms of the original variables (\ref{st}) (or (\ref{stferm}) in the
presence of fermions) include the global scale transformations of the
metric,
of the scalar fields $\varphi_{a}$ related to the measure $\Phi$ (and
of the fermion fields) and in addition the dilaton field $\phi$ undergoes
a global shift. In the CEF  (see Eqs. (\ref{ct}) or 
(\ref{ctferm}) where
the theory is
formulated in the Riemannian (or Einstein-Cartan) space-time), all
dynamical variables are invariant under the transformations (\ref{st}) (or
(\ref{stferm})) except for the dilaton field which still gets shifted by a
constant. Thus, SSB of the scale symmetry that appears firstly in
(\ref{varphi}) when solving Eq. (\ref{7}), is reduced, in the CEF,
 to SSB of the shift symmetry   $\phi\rightarrow\phi
+const$.   

The original action does not includes potentials but in the
CEF, the exponential potential
appears as a result of SSB of the scale symmetry. In the generic case
$\Delta
=b_{g}-b_{k}\neq
0$, the process of SSB also produces terms with 
higher powers in derivatives of the dilaton field.

Cosmological scaling solutions of the theory were studied.
The flatness of the potential $V_{eff}$ which is associated here with the
exponential form, is protected by the scale symmetry.
 Quintessence
solutions (corresponding to accelerating universe) were found possible
 for a broad range of parameters.

Finally, the behavior of fermions in such type of models was investigated.
Scale invariant fermion mass-like terms can be introduced in two different
ways since they can appear coupled to each of the two different measures
of the theory. Although an exponential of the dilaton field $\phi$
couples to the fermion in both of these terms, it is found that when the
fermions are treated as a test particles in the scaling background, their
masses in the CEF are constants.

Even more surprising is the behavior of the fermions in the limit of high
fermion density as compared to the dilaton density. This approximation is
regarded as more realistic if we are interested in the regular particle
physics behavior of these fermions under normal laboratory conditions.
It is found then that in the CEF, a given fermion can
behave in three different ways
according to the three different solutions of the fundamental constraint
(\ref{confermEin}). Two of the solutions correspond to fermions with
constant masses and the other - to a higher fermion self-interaction
which, we expect, can generate mass on the quantum level in a manner
similar
to a NJL model\cite{NJL}. From one primordial fermion three are
obtained for free. This suggests a new approach to the "family problem" in
particle physics.

In addition to this, for the two mentioned above solutions (\ref{zeta12}) 
corresponding to constant fermion masses, the fermion-dilaton coupling 
in the CEF (proportional to $F(\zeta)$,
Eq.(\ref{F})) disappears automatically. If one of these types of fermions
is associated to the first family (regular matter, i.e., $u$ and $d$
quarks, $e^{-}$ and $\nu_{e}$), we obtain that normal matter decouples
from the dilaton.

All what has been done here concerning fermions is in the context of a toy
model without Higgs fields, gauge bosons and the associated
$SU(2)\times U(1)\times SU(3)$ gauge symmetry of the standard model. As we
have seen
in other models (see \cite{GK4}, the second reference of \cite{G1} and
\cite{K}), it is possible to incorporate the two measure ideas  with the
gauge symmetry and Higgs mechanism. Now the differences consist of:
i) the presence of global scale symmetry, ii) the most general TMT 
structure for gravitation and dilaton sector.                                 
The complete discussion of the standard model in the context of such TMT
structure will be presented in a separate publication\cite{prep}.
Here we want only to explain shortly the main ideas that provides us the
possibility to implement this program.
It is important that in a simple way gauge fields can be incorporated so
that they will not appear in the fundamental constraint\footnote{This may
be 
done by making the gauge field kinetic terms coupled only to $\sqrt{-g}$ 
which is dictated  by local scale invariance of that part of the 
action.}
in contrast to the fermions (see for comparison Eq. (\ref{confermEin})).
We can also work
without significant changes in the discussion of the fermionic sector if
instead of explicit mass-like terms we will work with similar terms where
the coupling constants with the dimensionality of the mass   are replaced
by
gauge invariant Yukawa couplings to the Higgs field. Proceeding in the
spontaneously broken $SU(2)\times U(1)$ gauge theory\cite{prep} and
starting from one  correspondent primordial fermions family
we observe again\cite{prep} the effect of generation of
three fermion families, as was above in the toy model. Generating
mass of two of them is automatic as in the previous discussion. For the
third we need again some quantum effect that gives rise to a fermion
condensate.

The analysis of the constraint (\ref{confermEin}) 
provides in general seven solutions for $\zeta$. It could be that among of
them there is a solution corresponding to a fermionic state responsible
for dark matter. For example, the solution (\ref{srtzeta}) after inserting
into $00$-component of the energy-momentum tensor (\ref{Tmn+f}) makes
the last three
terms of (\ref{Tmn+f}) to be dependent in the same manner
only on the  combination $\overline{\Psi}^{\prime}\Psi^{\prime}
e^{\frac{1}{2}\alpha\phi /M_{p}}$ and they appear to be of the same order
of magnitude.
This
implies that
fermion contributions to the energy and those of
the scalar field are of the same order that provides then a possible
explanation of the "cosmic coincidence" problem. A consistent study of
these cosmological questions will become possible after we will explore in
detail\cite{prep} the field theoretic aspects of the displayed here
"families birth effect".

\section{Acknowledgments}

We are grateful to
J. Alfaro, S. de Alwis, M. Banados, J. Bekenstein,  Z. Bern, R. Brustein,  
K. Bronnikov, V. Burdyuzha,
L. Cabral, S. del Campo, C. Castro, A. Chernin,
A. Davidson, G. Dvali, M. Eides,  P. Gaete, H. Goldberg, M. Giovannini, A.
Guth, 
F. Hehl,
V. Ivashchuk, V. Kiselev, A. Linde,
L. Horwitz, M. Loewe, A. Mayo, Y. Ne'eman, Y. Jack Ng,
M. Pavsic, M.B.Paranjape, L. Parker,
J. Portnoy, V.A. Rubakov,
E. Spallucci, Y. Verbin, A. Vilenkin, M. Visser, K. Wali, C. Wetterich, 
P. Wesson
and J.Zanelli for discussions on different aspects of this paper.

\end{document}